\documentclass[slac_one]{revtex4}
\usepackage{graphicx}
\usepackage{fancyhdr}
\usepackage{amssymb} 
\def\DAF{DA\char8NE}
\def\ifm#1{\relax\ifmmode#1\else$#1$\fi} \def\epm{\ifm{e^+e^-}}
\newcommand{\cm}{{\rm \,cm}}
\newcommand{\eV}{{e\kern-.07em V}}
\newcommand{\MeV}{{\rm \,M\eV}}

\newcommand{\kl}{\mbox{$K_L$}}
\newcommand{\ks}{\mbox{$K_S$}}
\newcommand{\kcr}{\ensuremath{K_\mathrm{crash}}}
\newcommand{\DKSee}{\ensuremath{K_S\rightarrow e^+e^-}}
\newcommand{\DKLee}{\ensuremath{K_L\rightarrow e^+e^-}}

\newcommand{\DKLmm}{\ensuremath{K_L\rightarrow \mu^+\mu^-}}
\newcommand{\DKSpippim}{\ensuremath{K_S\rightarrow\pi^+\pi^-}}
\newcommand{\Dphipippimpio}{\ensuremath{\phi\rightarrow\pi^+\pi^-\pi^0}}
\newcommand{\AmS}{{\protect\the\textfont2
  A\kern-.1667em\lower.5ex\hbox{M}\kern-.125emS}}
\newcommand{\ke}{\ensuremath{K^\pm\to e^\pm\nu_e}}
\newcommand{\km}{\ensuremath{K^\pm\to \mu^\pm\nu_\mu}}
\def\Journal#1#2#3#4{{\it #1} {\bf #2}, #3 (#4)}

\begin{document}

\title{K$^\pm$e2 search and Lepton Flavor Violation at KLOE} 
%

\author{B. Sciascia, on behalf of the KLOE Collaboration 
\footnotetext[1]{KLOE Collaboration:
F.~Ambrosino,
A.~Antonelli,
M.~Antonelli,
F.~Archilli,
P.~Beltrame,
G.~Bencivenni,
S.~Bertolucci,
C.~Bini,
C.~Bloise,
S.~Bocchetta,
F.~Bossi,
P.~Branchini,
P.~Campana,
G.~Capon,
T.~Capussela,
F.~Ceradini,
F.~Cesario,
P.~Ciambrone,
F.~Crucianelli,
E.~De~Lucia,
A.~De~Santis,
P.~De~Simone,
G.~De~Zorzi,
A.~Denig,
A.~Di~Domenico,
C.~Di~Donato,
B.~Di~Micco,
M.~Dreucci,
G.~Felici,
M.~L.~Ferrer,
S.~Fiore,
P.~Franzini,
C.~Gatti,
P.~Gauzzi,
S.~Giovannella,
E.~Graziani,
W.~Kluge,
V.~Kulikov,
G.~Lanfranchi,
J.~Lee-Franzini,
D.~Leone,
M.~Martini,
P.~Massarotti,
S.~Meola,
S.~Miscetti,
M.~Moulson,
S.~M\"uller,
F.~Murtas,
M.~Napolitano,
F.~Nguyen,
M.~Palutan,
E.~Pasqualucci,
A.~Passeri,
V.~Patera,
F.~Perfetto,
P.~Santangelo,
B.~Sciascia,
A.~Sciubba,
A.~Sibidanov,
T.~Spadaro,
M.~Testa,
L.~Tortora,
P.~Valente,
G.~Venanzoni,
R.Versaci}}
\affiliation{Laboratori Nazionali di Frascati dell'INFN}

\begin{abstract}
This paper is devoted to the first analyses based on the complete data sample
collected by the KLOE detector at \DAF, the Frascati $\phi$-factory.
The result for the BR(K$_S \to \gamma\gamma)$ and the search for the
decay K$_S \to e^+ e^-$ are presented. Particular emphasis is put on the 
measurement of the ratio of $K_{e2}$ and $K_{\mu2}$ BR's.
\end{abstract}

\maketitle

\thispagestyle{fancy}


\section{EXPERIMENTAL SETUP}
DA$\Phi$NE, the Frascati $\phi$ factory, is an $e^{+}e^{-}$ collider
working at $\sqrt{s}\sim m_{\phi} \sim 1.02$~GeV. $\phi$ mesons are produced 
nearly at rest, with a visible cross section of $\sim$~3.1~$\mu$b
and decay into $\ks\kl$ (BR$\sim 34$\%) or $K^+K^-$ (BR$\sim 49$\%);
Neutral and charged kaons have momenta of 110 and 127 \MeV, respectively.

The kaon pairs from $\phi$ decay are produced in a pure $J^{PC}=1^{--}$ 
quantum state, so that 
the detection of a \ks(\kl) thus signals, or tags, the presence of a \kl(\ks).
This in effect creates pure \ks\ and \kl\ beams of precisely known momenta
(event by event, from kinematic closure) and flux, which can be used to
measure absolute \ks\ and \kl\ BRs. Similar arguments hold for $K^+$ and $K^-$ as well.
$K_S$ and $K_L$ can be distinguished by their mean decay lengths:
$\lambda_{S} \sim $~0.6~cm and $\lambda_{L} \sim $~340~cm.

The analysis of kaon decays is performed with the KLOE detector~\cite{kloe}, 
consisting essentially of a drift chamber, DCH, surrounded by an
electromagnetic calorimeter, EMC. A superconducting coil provides a 0.52~T magnetic field.
The DCH 
is a cylinder of 4~m in diameter
and 3.3~m in length, which constitutes a fiducial volume 
for $K^\pm$ decays extending for $\sim1\lambda_\pm$, respectively.
The momentum resolution for tracks 
at large polar angle is $\sigma_{p}/p \leq 0.4$\%. 
The EMC is a lead/scintillating-fiber sampling calorimeter  
consisting of a barrel and two endcaps, with good
energy resolution, $\sigma_{E}/E \sim 5.7\%/\sqrt{\rm{E(GeV)}}$, and excellent 
time resolution, $\sigma_{T} =$~54~ps$/\sqrt{\rm{E(GeV)}} \oplus 50$ ps. 

In KLOE, the identification of \kl-interaction in the EMC (\kcr\ events in the following)
is used to tag the presence of \ks\ mesons.
$K^+$ and $K^-$ decay with a mean length of $\lambda_\pm\sim $~90~cm and can be 
distinguished from their decays in flight to one of the two-body final states 
$\mu\nu$ or $\pi\pi^0$.
The c.m. momenta reconstructed from identification of 1-prong 
$K^\pm\to\mu\nu,\pi\pi^0$ decay vertices in the DC 
peak around the expected values with a resolution of 1--1.5~MeV, 
thus allowing clean and efficient tagging. 

In early 2006, the KLOE experiment completed data taking, having collected
$\sim2.5$~fb$^{-1}$ of integrated luminosity at the $\phi$ peak,
corresponding to $\sim$3.8 billion $K^+K^-$ pairs, 
and to $\sim$2.6 billion $K_LK_S$ pairs

\section{MEASUREMENT OF BR(K$_S \to \gamma\gamma)$}

In ChPT calculations of the amplitude for K$_S \to \gamma\gamma$ process, since all particles
involved are neutral, there are non tree-level contributions. Moreover, at $\mathcal O(p^4)$,
only finite chiral-meson loops contribute. BR(K$_S \to \gamma\gamma)$) is predicted 
unambiguously at this level in terms of the couplings G$_8$ and G$_{27}$, giving
$2.1\times10^{-6}$~\cite{ksggth}. The most precise published measurement of this BR
is from NA48: BR = 2.78(6)(4)$\times10^{-6}$~\cite{ksggNA48}. 
This result would suggest the need for
a significant $\mathcal O(p^6)$ correction in the ChPT calculation of the BR.

\begin{figure*}[th]
\centering
\includegraphics[width=80mm]{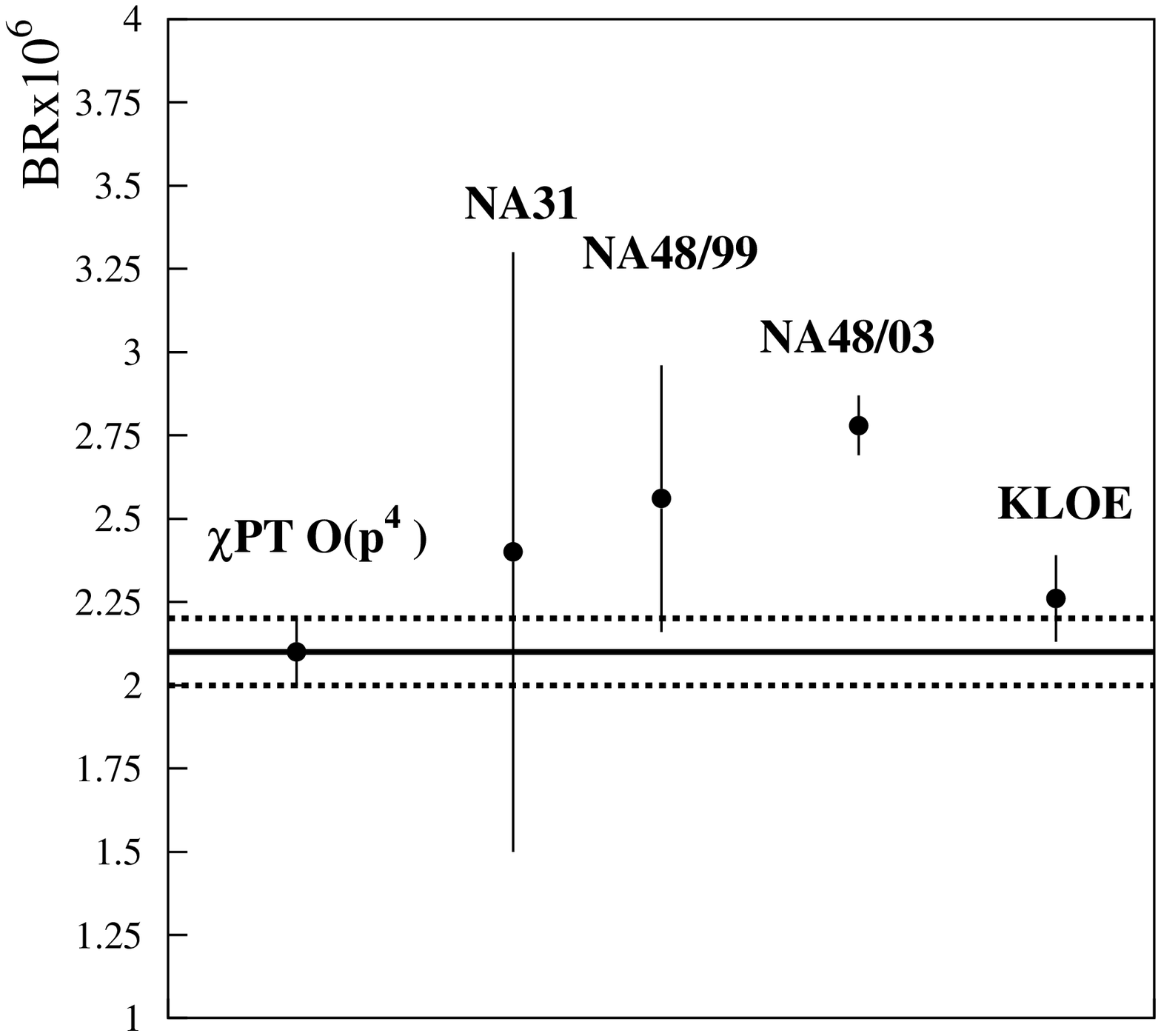}
\includegraphics[width=83mm]{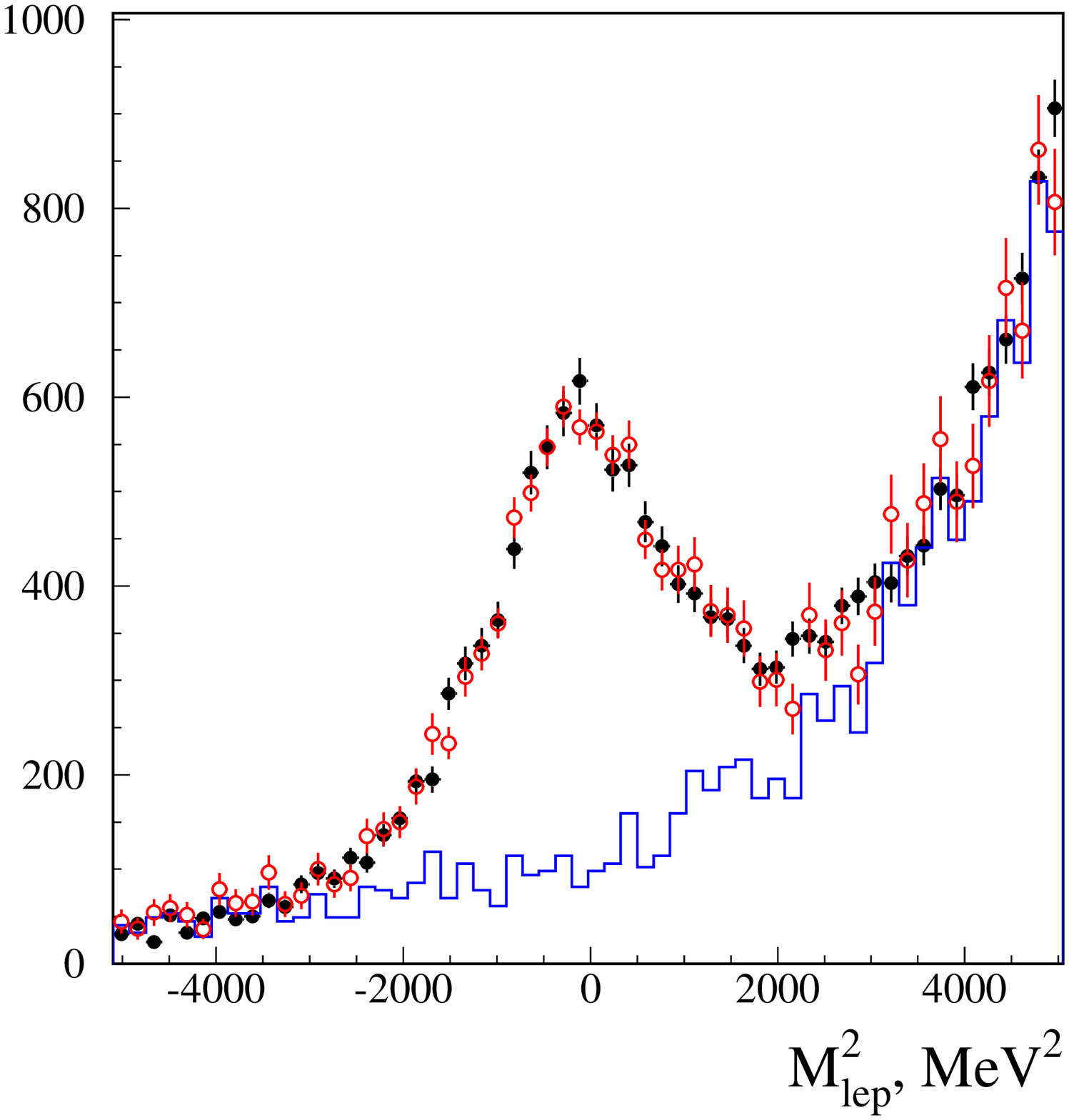}
\caption{{\bf left:} Comparison of BR$(K_S \to \gamma \gamma)$ measurements 
and ChPT predictions. {\bf right:} Distributions of the lepton mass squared $M^2_\mathrm{lep}$
of the secondary track for \ke\ and \km\ events. 
Filled dots represent the data, open dots
are the result of a maximum-likelihood fit using signal and
background (solid line) distributions as input. } \label{ksggres}
\end{figure*}

KLOE searched for the decay $K_S \to \gamma\gamma$
in a sample of $\sim 2 \times 10^9$ $\phi \to K_S K_L$
decays which corresponds to an integrated luminosity of 1.9 fb$^{-1}$.
Two prompt photons must be detected, and $K_S \to 2 \pi^0$ decays
counted in the same sample, are used as normalization sample.
KLOE measured~\cite{ksggpaper}: 
${\rm BR} (K_S \to \gamma\gamma)=(2.26\pm 0.12_{\rm stat}\pm 0.06_{\rm syst} ) \times 10^{-6}$.
This result deviates by 3 $\sigma$'s  from the previous
best determination, as shown in Fig.~\ref{ksggres}, left panel.
While the number of K$_S\to\gamma\gamma$ observed by KLOE is $\sim$700, as compared to
the $\sim$7500 observed by NA48, KLOE profits from the use of a tagged K$_S$ beam and
does not have to contend with irreducible background from K$_L\to\gamma\gamma$

Precise ChPT theory calculation for this decay are done
at $\mathcal O(p^4)$. Higher order effects are predicted
to be at most of the order of $\sim$ 20\% of the
$\mathcal O(p^4)$  decay amplitude. Our measurement is consistent
with negligible higher order corrections.

\section{SEARCH FOR THE DECAY K$_S \to e^+ e^-$}
The decay \DKSee, like the decay  \DKLee\ or \DKLmm, is a flavour-changing 
neutral-current process, suppressed in the Standard Model and dominated by 
the two-photon intermediate state~\cite{kseetheory}.
For both \ks\ and\kl, the \epm\ channel is much more suppressed than the 
$\mu^+\mu^-$ one (by a factor of $\sim 250$). 
Using Chiral Perturbation Theory  ($\chi_{PT}$) to order $\mathcal{O}(p^4)$, 
the SM prediction for BR$(\DKSee)$
is evaluated to be  $\sim 2 \times 10 ^{-14}$.
A value significantly higher than expected would point to new physics.
The best experimental limit for $BR(\DKSee)$ has been measured
by CPLEAR~\cite{cplearksee}, and it is equal to $1.4\times 10^{-7}$, at $90\%$~CL.

$\sim 650$ million \kcr\ events are used as a starting sample for the \DKSee\ search. 
\DKSee\ events are selected by requiring the presence of two tracks of opposite 
charge with their point of closest approach to the origin inside a cylinder $4\cm$ 
in radius and $10\cm$ in length along the beam line. 
The track momenta and polar angles must satisfy the fiducial cuts 
$120\le p \le 350\MeV$ and $30^{\circ} \le \theta \le 150^{\circ}$. 
The tracks must also reach the EMC without spiralling, and have an associated 
cluster.  

The two-track invariant mass is evaluated in electron hypothesis ($M_{ee}$). 
A preselection cut requiring $M_{ee}> 420 \MeV$ has been applied, which rejects most
of $\DKSpippim$ events, for which $M_{ee}\sim 409 \MeV$. The residual background has 
two main components: \DKSpippim\ events, populating the low $M_{ee}$ region, and 
\Dphipippimpio\ events, spreading over the whole spectrum. The \DKSpippim\ events
have such a wrong reconstructed $M_{ee}$ because of track resolution or one pion 
decaying 
into a muon. The \Dphipippimpio\ events enter the preselection because of a machine 
background cluster, accidentally satisfying the \kcr\ algorithm. 
After preselection we are left with $\sim 5\times 10^5$ events.
To have a better separation between signal and background, a $\chi^2$-like 
variable is defined, collecting information from the clusters associated
to the candidate electron tracks.
A signal box to select the \DKSee\ events can be conveniently defined in the 
$M_{ee}-\chi^2$ plane.

The $\chi^2$ cut for the signal box definition has been chosen 
to remove all MC background events: $\chi^2 < 70$.
The cut on $M_{ee}$ is practically set by the $p^*_\pi$ cut,
which rules out all signal events with a radiated photon 
with energy greater than $20\MeV$, corresponding to an invariant mass window:
$477 <  M_{ee} \le 510 \MeV$.
The signal box selection on data gives $N_{obs}=0$. 
The upper limit at $90\%$ CL on the expected number of signal
events is $UL(\mu_S) = 2.3$.

The total selection efficiency on \DKSee\ events is evaluated by MC, 
and includes contribution from radiative corrections. 
The number of \DKSpippim\ events $N_{\pi^+\pi^-}$ counted on the same sample 
of \ks\ tagged events is used as normalization.
The upper limit on BR(\DKSee) is evaluated as follows:
\begin{equation}
 UL(BR(\DKSee)) = 
 UL(\mu_s)     \times 
\frac{\epsilon_{\pi^+\pi^-}(sele \vert K_{crash})}{
\epsilon_{sig}(sele \vert K_{crash})}\times
\frac{BR(\DKSpippim)}{N_{\pi^+\pi^-}}. \nonumber 
\end{equation}
Using $\epsilon_{sig}(sele \vert K_{crash}) = 0.480(4)$,
$\epsilon_{\pi^+\pi^-}(sele \vert K_{crash}) = 0.6102(5)$ and 
$N_{\pi^+\pi^-} = 217,422,768$, we obtain
\begin{equation}
UL(BR(\DKSee(\gamma))) =  9 \times 10^{-9}, \; {\rm at} \;90\%\,{\rm CL}\, .
\end{equation}
Our measurement improves by a factor of $\sim 15$ on the
CPLEAR result~\cite{cplearksee}, for the first time including radiative
corrections in the evaluation of the upper limit.

\section{MEASUREMENT OF R$_K$}

A strong interest for a new measurement of the ratio
$R_K=\Gamma(\ke)/\Gamma(\km)$
has recently arisen, triggered by the work of Ref.~\cite{masiero}. 
The SM prediction of $R_K$ benefits from
cancellation of hadronic uncertainties to a large extent and therefore
can be calculated with high precision. Including radiative
corrections, the total uncertainty is less than 0.5 per
mil~\cite{ciriglianorosell07}. Since the electronic channel is
helicity-suppressed by the $V-A$ structure of the charged weak
current, $R_K$ can receive contributions from physics beyond the SM,
for example from multi-Higgs effects inducing an effective
pseudoscalar interaction.  It has been shown in Ref.~\cite{masiero}
that deviations from the SM of up to few percent on $R_K$ are quite
possible in minimal supersymmetric extensions of the SM and in
particular should be dominated by lepton-flavor violating
contributions with tauonic neutrinos emitted. Using the present KLOE
dataset of $\sim$2.5 fb$^{-1}$ of luminosity integrated at the
$\phi$-meson peak, we show that an accuracy of about 1~\% in the
measurement of $R_K$ might be reached.

In order to compare with the SM prediction at this level of accuracy,
one has to treat carefully the effect of radiative corrections, which
contribute several percent to the $K_{e2}$ width.  In particular, the
SM prediction of Ref.~\cite{ciriglianorosell07} is made considering all
photons emitted by the process of internal bremsstrahlung (IB) while
ignoring any contribution from structure-dependent direct emission
(DE).  Of course both processes contribute, so in the analysis we will
consider DE as a background which can be distinguished from the IB
width by means of a different photon energy spectrum.

Given the $K^\pm$ decay length of $\sim$90~cm, the selection of
one-prong $K^{\pm}$ decays in the DC required to tag $K^{\mp}$ has an
efficiency smaller than 50\%. In order to keep the statistical
uncertainty on the number of \ke\ counts below 1\%, we decided to
perform a ``direct search'' for \ke\ and \km\ decays, without
tagging. Since we measure a ratio of BR's for two channels with
similar topology and kinematics, we expect to benefit from some
cancellation of the uncertainties on tracking, vertexing, and
kinematic identification efficiencies.  Small deviations in the
efficiency due to the different masses of $e$'s and $\mu$'s can be
evaluated using MC.

A powerful kinematic variable used to distinguish \ke\ and \km\ decays
from the background is calculated from the momenta of the kaon and the
secondary particle measured in DC: assuming zero neutrino mass one can
obtain the squared mass of the secondary particle, or lepton mass
($M_\mathrm{lep}^2$). While the one-prong selection is enough for clean
identification of a \km\ sample, further rejection is needed in order
to identify \ke\ events: the background, which is dominated by badly
reconstructed \km\ events, is reduced by a factor of $\sim$10 by the
quality cuts, but still remains $\sim$10 times more frequent than the
signal in the region around the electron mass peak. 
Information from the EMC is used to improve background rejection:
Electron clusters can be further distinguished from $\mu$ (or $\pi$)
clusters by exploiting the granularity of the EMC, in particular using
the spread of energy deposits on each plane ($E_\mathrm{RMS}$).
The PID technique described above selects \ke\ events with an
efficiency $\epsilon^\mathrm{PID}_{Ke2}\sim64.7(6)\%$ and a rejection
power for background of $\sim300$. These numbers have been evaluated
from MC.
A likelihood fit to the two-dimensional $E_\mathrm{RMS}$ vs
$M_\mathrm{lep}^2$ distribution was performed to get the number of
signal events. Distribution shapes for signal and background were
taken from MC; the normalizations for the two components are the only
fit parameters.  The number of signal events obtained from the fit is
$N_{Ke2}=8090\pm156$. Projections of the fit results onto the $M^2_\mathrm{lep}$ axes
is compared to real data in Fig.~\ref{ksggres}, right panel.

The primary generators for \ke\ and \km\ decays include radiative
corrections and allow for the emission of a single photon in the final
state~\cite{radcor}.  $\ke+\gamma$ events with photon energy in the
kaon rest frame $E_{\gamma}<20$~MeV (where the DE contribution is indeed negligible) 
were considered as signal.

The number of \km\ events in the same data set is extracted from a
similar fit to the $M_\mathrm{lep}^2$ distribution
The fraction of background events under the
muon peak is estimated from MC to be less than one per mil.  The
number of \km\ events is 499\,251\,584$\pm$35403.
Using the number of observed \ke\ and \km\ events and all
corrections, we get the preliminary result~\cite{pos:ke2}
\begin{equation}
\label{eq:preliminary}
R_K = (2.55\pm0.05\pm0.05)\times10^{-5}.
\end{equation}
This value is compatible within the error with the SM prediction, $R_K
= (2.477\pm0.001)\times10^{-5},$ and with other recent
measurements by NA48~\cite{NA48}.

Three sources contribute to the present statistical uncertainty of
1.9\%: fluctuation in the signal counts (1.1\%), fluctuation in the
background to be subtracted (0.7\%), and statistical error on the MC
estimate of the background (1.4\%).
The total error on $R_K$ should be reduced to $\sim1.3\%$ after 
analysis completion.


\end{document}